\documentclass[prb,twocolumn,showpacs,preprintnumbers,amsmath,amssymb,floatfix]{revtex4}

\usepackage{graphicx,float}
\usepackage{subfigure}
\usepackage{dcolumn}
\usepackage{bm}
\usepackage{mathrsfs}
\usepackage{color}

\begin{document}

\title{Robust creation of entangled states of two coupled flux qubits\\via dynamic control of the transition frequencies}

\author{Keyu \surname{Xia}}
\email{Keyu.Xia@mpi-hd.mpg.de}
\affiliation{Max-Planck-Institut f\"{u}r Kernphysik, Saupfercheckweg 1,
D-69117 Heidelberg, Germany}

\author{Mihai \surname{Macovei}}
\affiliation{Max-Planck-Institut f\"{u}r Kernphysik, Saupfercheckweg 1,
D-69117 Heidelberg, Germany}

\author{J\"{o}rg \surname{Evers}}%
\email{joerg.evers@mpi-hd.mpg.de}
\affiliation{Max-Planck-Institut f\"{u}r Kernphysik, Saupfercheckweg 1,
D-69117 Heidelberg, Germany}

\author{Christoph H. \surname{Keitel}}
\affiliation{Max-Planck-Institut f\"{u}r Kernphysik, Saupfercheckweg 1,
D-69117 Heidelberg, Germany}

\date{\today}

\begin{abstract}
Coherent control and the creation of entangled states are discussed
in a system of two superconducting flux qubits interacting with each
other through their mutual inductance and identically coupling to a
reservoir of harmonic oscillators. We present different schemes
using continuous-wave control fields or Stark-chirped rapid
adiabatic passages, both of which rely on a dynamic control of the
qubit transition frequencies via the external bias flux in order to
maximize the fidelity of the target states. For comparison, also
special area pulse schemes are discussed. The qubits are operated 
around the optimum point, and decoherence is modelled via a bath of 
harmonic oscillators. As our
main result, we achieve controlled robust creation of different Bell
states consisting of the collective ground and excited state of the
two-qubit system.
\end{abstract}


\maketitle


\section{Introduction}
Superconducting solid-state qubits are promising candidates for quantum computation and quantum information, for example, because of their inherent scalability using well-established micro fabrication techniques and the ability to design them to meet specific characteristics~\cite{JClarkeNature453p1031,JMajerNature449p443}.
The entanglement of multiple qubits is at the heart of quantum computation and quantum information, and thus has been studied extensively in the past few years. For example, entanglement between a superconducting flux qubit and a quantum harmonic oscillator such as a superconducting quantum interference device (SQUID)~\cite{IChiorescuNature431p159} or a LC circuit~\cite{JJohanssonPRL96p127006}  were examined, as well as between two qubits \cite{JClarkeNature453p1031}.
Entanglement or coupling between two superconducting charge or flux qubits are well studied in theory~\cite{YXLiuPRB76p144518,PBertetPRB73p064512} and also observed in experiments~\cite{TYamamotoPRB77p064505,JBMajerPRL94p090501,AIzmalkovPRL93p037003,RMcdermottScience307p1299}. Using Josephson charge qubits coupled by an inductance, a scalable quantum computing architecture was proposed~\cite{JQYouPRL89p197902}. Recently, a superconducting quantum device consisting of four coupled flux qubits was achieved experimentally~\cite{MGrajcarPRL96p047006}.

A key limiting factor for all of these devices is decoherence such as dephasing or energy relaxation, which occurs, e.g., when the quantum devices couple to environmental degrees of freedom at a finite temperature. For example, charge qubits are very sensitive to background charge fluctuations~\cite{JClarkeNature453p1031,JQYouPhysToday58p42}. Flux qubits are practically insensitive to background charge fluctuations, but their phase coherence can still be destroyed by a large number of effects~\cite{YMakhlinRMP73p357}. Therefore, although superconducting qubits with long decoherence times up to the order of $\mu s$ are  available~\cite{JClarkeNature453p1031,JQYouPhysToday58p42,YYuPRL92p117904, EIlichevPRL91p097906,SHanScience293p1457},  applications are severely limited by the relaxation time. Zhang et al. proposed a method to protect stationary entanglement in superconducting qubits from the relaxation and dephasing processes \cite{JZhangarxiv0808}. They found a maximum stationary concurrence of about $1/3$ and fidelity of $2/3$.

Another issue is the realization of robust quantum operations. To date, operations on SQs are typically based on special-area pulses~\cite{RMcdermottScience307p1299,YXLiuPRL96p067003,JMajerNature449p443,MASillanpaaNature449p438}, which require an accurate control of field parameters such as intensity, duration and shape. In a recent experiment, pulse-timing uncertainty limited the achieved fidelity by an error of about $10\%$~\cite{JClarkeNature453p1031,MSteffenPRL97p050502}. Similar problems are faced in the preparation of atomic systems, where coherent population transfer schemes have been developed in order to overcome these limitations~\cite{RevModPhys.70.1003,coherent-passage}. In contrast to atoms, superconducting qubits have the advantage that the transition frequencies can be changed to a large degree on demand via the bias flux~\cite{JMajerNature449p443,AIzmalkovPRL93p037003,MGrajcarPRL96p047006}, bias charge~\cite{JClarkeNature453p1031} or bias current \cite{LFWeiPRL100p113601}. Making use of this advantage, very recently, based on the breaking of parity symmetries, a coherent population transfer in superconducting current-biased phase qubit was demonstrated using so-called Stark-chirped rapid adiabatic passages (SCRAPs)~\cite{coherent-passage,LFWeiPRL100p113601,ZFicekQuantInterfCoh2004}.

Here, we discuss schemes to create a set of relevant collective states in a system of two inductively coupled flux qubits driven by time-dependent magnetic fluxes (TDMFs), addressing both the questions of robust coherent control and of decoherence. Controlled state transfer is reported using three different techniques: Continuous-wave control fields, robust SCRAP-based state transfer, and special-area pulses. A dynamic control of the qubit transition frequencies via the external bias flux is applied in order to maximize the fidelity of our target states.  The inevitable decoherence is modelled via an interaction with a reservoir consisting of an ensemble of harmonic oscillators. Our approach has the advantage that the qubits operate around the optimum point, where the energy levels are symmetric as a function of the bias flux. In addition, flux qubits typically feature longer decoherence times compared to current-biased qubits~\cite{JQYouPhysToday58p42,YYuPRL92p117904}.

In particular, we first demonstrate how  continuous-wave driving fields can be used to populate the antisymmetric collective state for a pair of identical qubits, which is usually decoupled from external  electromagnetic field and thus hard to populate. Second, we demonstrate efficient population of the collective symmetric state, both directly and via the anti-symmetric state. Finally,  we discuss controlled population of the collective excited state and of different Bell states composed of the collective ground and excited states using SCRAP. Interestingly, in contrast to atomic systems, the population of the collective excited state from the ground state is possible directly without involving the intermediate symmetric and anti-symmetric states.

The outline is as follows: In Sec.~\ref{sec:Model}, we introduce the Hamiltonian of our system, which is general in the sense that different mutually exclusive processes are modelled, which become relevant depending on the parameters of the applied fields. This Hamiltonian will successively be used throughout the different control schemes presented. In Sec.~\ref{sec:g2a}, the antisymmetric state is populated with a near unit population for two identical flux qubits. Sec.~\ref{sec:a2s} studies the preparation of the symmetric state both from the ground and from the antisymmetric state. In Section~\ref{sec:g2e}, we propose a robust method to create the collective excited state and superpositions of different Bell states. Finally, Section~\ref{sec:conclusion} briefly discussed and summarizes the results.


\section{\label{sec:Model}Model}
\begin{figure}[t]
\centering
\includegraphics[width=0.6\linewidth]{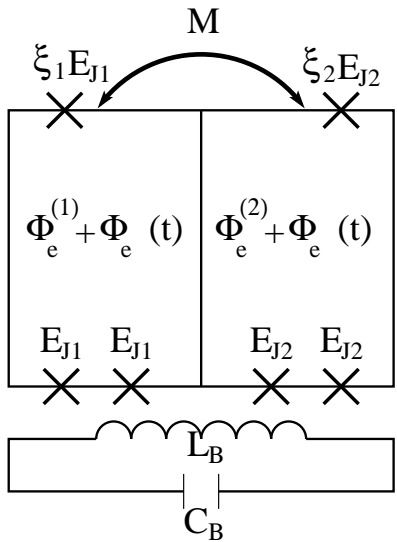}
\caption{\label{fig:system}Two superconducting flux qubits interacting with each other through their mutual inductance $M$ and damped to a common reservoir modeled as an LC circuit. The individual bias fluxes are varied dynamically in order to control the qubit transition frequencies around the optimum point. In the figure, crosses indicate Josephson junctions, whereas the bottom circuit loop visualizes the bath.}
\end{figure}

 Our model system consists of two flux qubits coupled to each other through their mutual inductance $M$~\cite{YXLiuPRL96p067003} and to a reservoir of harmonic oscillators modeled as an LC circuit, see Fig.~\ref{fig:system}.
Each qubit loop contains three junctions: two identical ones and one which is smaller by a factor of $\xi_{l}$. The Josephson energies and capacitances in the $l$th qubit loop are given by  ($l\in\{1,2\}$)
\begin{subequations}
\begin{align}
E_{J1}^{(l)}&=E_{J2}^{(l)}=E_{J}^{(l)},
\qquad E_{J3}^{(l)}=\xi_l E_{J}^{(l)} \,,\\
C_{J1}^{(l)}&=C_{J2}^{(l)}=C_{J}^{(l)}
, \qquad  C_{J3}^{(l)}=\xi_l C_{J}^{(l)}\,.
\end{align}
\end{subequations}
 The gauge-invariant phase drops across the three junctions in the $l$th qubit are $\phi_{1}^{(l)},\phi_{2}^{(l)}$ and $\phi_{3}^{(l)}$.
 Both qubits experience the same TDMF
\begin{align}
\Phi_e(t)=A\, \cos(\omega_c t)\,,
\end{align}
but an individual bias magnetic flux $\Phi_e^{(l)}$ is applied through each qubit.

The qubits can be visualized as artificial two-level atoms coupling to a common reservoir of quantum oscillators.
Using the phase constraint condition through the $l$th qubit loop \begin{align}
\sum_{i=1}^{3} \phi_i^{(l)}+\left (2\pi \Phi_e^{(l)}/\Phi_0 \right )+2\pi \Phi_e (t)/\Phi_0=0\,,
\end{align}
where $\Phi_0=h/(2e)$ is the flux quantum, the system in Fig.~\ref{fig:system} can be described by the total Hamiltonian $H=H_Q+H_B$. The two-qubit Hamiltonian $H_Q$ in two-level approximation and rotating wave approximation is given by~\cite{YXLiuPRL96p067003}
\begin{align}\label{eq:HQ}
    H_Q& =\frac{1}{2} \sum_{l=1}^2 {\hbar \omega_0^{(l)} \sigma_z^{(l)}} \nonumber\\
     & \quad -\hbar\sum_{l=1}\left (k_l \sigma_+^{(l)}e^{-i\omega_ct}+H.c. \right )\nonumber\\
     & \quad -\hbar\sum_{l\neq m=1}^2
     \left (\Omega_{lm}^{(1)}\sigma_+^{(l)}\sigma_-^{(m)}+H.c. \right )
\left (e^{i\omega_c t}+e^{-i\omega_c t} \right )\nonumber\\
     & \quad -\hbar\sum_{l\neq m=1}^2
     \left (\Omega_{lm}^{(2)}\sigma_+^{(l)}\sigma_+^{(m)}e^{-i\omega_c t}+H.c. \right )\nonumber\\
     & \quad +\hbar \left (\lambda_1\sigma_+^{(1)}\sigma_-^{(2)}+\lambda_2\sigma_+^{(1)}\sigma_+^{(2)}+H.c. \right )\,.
\end{align}
The transition frequency $\omega_0^{(l)}$ of qubit $l$ is determined by $\hbar\omega_0^{(l)}=\sqrt{t^2_l+\varepsilon_l^2}$ with the tunnel coupling $t_l$ between two wells in the $l$th qubit and the energy difference $\varepsilon_l$ of the wells measured with respect to the degeneracy point.  This frequency can be expressed as $\varepsilon_l=2I^{(l)}(\Phi_e^{(l)}-\Phi_0/2)$ with the persistent supercurrent $I^{(l)}$ and the bias flux $\Phi_e^{(l)}$ in the $l$th qubit loop \cite{JClarkeNature453p1031}.
The Pauli matrices of  the $l$th qubit with ground state $|g_l\rangle$ and excited state $|e_l\rangle$ are defined as
\begin{subequations}
\begin{align}
\sigma_z^{(l)}&=|e_l\rangle\langle e_l|-|g_l\rangle \langle g_l|\,,\\
\sigma_+^{(l)}&=|e_l\rangle \langle g_l|\,,\\ \sigma_-^{(l)}&=|g_l\rangle \langle e_l|\,.
\end{align}
\end{subequations}
The phases and amplitudes of the coupling strengths $k_l, \Omega_{lm}^{(1)}$ and $\Omega_{lm}^{(2)}$ can be controlled by the applied TDMFs. The always-on coupling parameters are given by~\cite{YXLiuPRL96p067003}
\begin{subequations}
\begin{align}
\hbar\lambda_1&=M \left \langle
e_1,g_2 \left |I^{(1)}I^{(2)} \right |g_1,e_2 \right \rangle\,,\\
\hbar\lambda_2&=M \left \langle e_1,e_2 \left |I^{(1)}I^{(2)} \right |g_1,g_2 \right\rangle\,.
\end{align}
\end{subequations}
Since $|\lambda_2|\ll \omega_l$, terms proportional to $\lambda_2$ and its complex conjugate can be neglected. Our system works near the optimal point
\begin{align}
f_l=\Phi_e^{(l)}/\Phi_0=1/2\,,
\end{align}
where the parameter $\lambda_1$ determined by the persistent supercurrent is a real number~\cite{YXLiuPRL96p067003,JQYouPRB71p024532}.

The interaction between flux qubits and the reservoir can be described by the Jaynes-Cummings Hamiltonian~\cite{ScZu1997,ZFicekQuantInterfCoh2004,YMakhlinRMP73p357,MASillanpaaNature449p438,AWallraffNature431p162,ABlaisPRA75p032329},
\begin{align}\label{eq:HB}
H_B& = \sum_r \hbar \omega_r a^{\dag}_r a_r -\hbar\sum_{l=1}^2\sigma_x^{(l)}
\sum_r \left(\eta_r^{(l)}a_r+ H.c. \right)\nonumber\\
& \quad -\hbar\sum_r\sum_{l\neq m=1}^{2} \left (\chi_r^{(lm)}\sigma_+^{(l)}\sigma_+^{(m)}a_r
+ H.c. \right )\,.
\end{align}
The bath oscillators have frequencies $\omega_r$, and $\langle a_r^\dag a_r \rangle=n_r$ is the average photon number of the $r$th field  mode. The thermal average number of photons in the oscillator
\begin{align}
N_{th}(\omega)=[\exp(\hbar \omega/k_B T)-1]^{-1}
\end{align}
is assumed negligible at the frequencies relevant to our system. $\eta_r^{(l)}$ and $\chi_r^{(lm)}$ are determined by the vacuum field.
In our model, we dropped the vacuum-induced decay with difference frequency $|\omega_0^{(1)}-\omega_0^{(2)}|$.
Since the magnitude of the vacuum field is proportional to the square root of its frequency \cite{RHLehmbergPRA2p883}, and because $|\omega_0^{(1)}-\omega_0^{(2)}| \ll \{\omega_0^{(1)}$, $\omega_0^{(2)}$, $\omega_0^{(1)}+\omega_0^{(2)}\}$, processes at the difference frequency are highly suppressed.

The diagonalization of the Hamiltonian in Eqs.~(\ref{eq:HQ}) and (\ref{eq:HB}) without TDMF leads to the eigenenergies $E_j$ and corresponding eigenstates $|j\rangle$  ($j\in\{e,s,a,g\}$)~\cite{ZFicekPhysRep372p369,ZFicekQuantInterfCoh2004}, which can be interpreted as a single four-level system,
\begin{subequations}
\label{eq:CBasis}
\begin{align}
|e\rangle &= |e_1,e_2\rangle\,,& E_e &=\hbar \omega_0\,, \\
|s\rangle &= \beta|e_1,g_2\rangle+\alpha|g_1,e_2\rangle\,,& E_s &=\hbar w\,, \\
|a\rangle &= \alpha|e_1,g_2\rangle-\beta|g_1,e_2\rangle\,,& E_a &=-\hbar w\,, \\
|g\rangle &= |g_1,g_2\rangle\,, & E_g &=-\hbar \omega_0\,,
\end{align}
\end{subequations}
with
\begin{subequations}
\begin{align}
w&=\sqrt{\Delta^2+\lambda^2}\,, &\qquad d&=\Delta+w\,,\\
\alpha&=\frac{d}{\sqrt{d^2+\lambda^2}}\,, & \qquad \beta&=\frac{\lambda}{\sqrt{d^2+\lambda^2}}\,,\\
\omega_0 &= \frac{\omega_0^{(1)}+\omega_0^{(2)}}{2}\,, & \qquad
\Delta &= \frac{\omega_0^{(2)}-\omega_0^{(1)}}{2}\,.
\end{align}
\end{subequations}
$\lambda=\lambda_1-\Omega_d$ includes the always-on coupling $\lambda_1$ and the shifts induced by the bath-induced dipole-dipole interaction (DDI) between the two qubits $\Omega_d$. For $\Delta=0$, the coefficients in Eq.~(\ref{eq:CBasis}) evaluate to $\alpha=\beta=1/\sqrt{2}$.

Applying the Born-Markov approximation to eliminate the bath \cite{ZFicekPhysRep372p369,ZFicekQuantInterfCoh2004,RHLehmbergPRA2p883}, we describe the bath-induced dissipation by the Liouville operators $\mathscr{L}\rho$ and $\mathscr{L}_\Gamma\rho$, where $\rho$ is the density matrix.
The first part $\mathscr{L}\rho$ results from the term proportional to $\eta_r^{(l)}$ in Eq.~(\ref{eq:HB}). It is composed of three
terms~\cite{ZFicekPhysRep372p369,ZFicekQuantInterfCoh2004}
\begin{equation}\label{eq:Lrho}
\mathscr{L}\rho= \left(\frac{\partial{\rho}}{\partial t}\right)_s+\left(\frac{\partial{\rho}}{\partial t}\right)_a+\left(\frac{\partial{\rho}}{\partial t}\right)_I,
\end{equation}
where
\begin{subequations}
\begin{align}
\left(\frac{\partial{\rho}}{\partial t}\right)_s &= -\Gamma_s \left \{(R_{ee}+R_{ss})\rho+\rho(R_{ee}+R_{ss}) \right . \nonumber \\
& \left . -2(R_{se}\rho R_{es}+R_{gs}\rho R_{sg}) \right \} \nonumber \\
& +(2\alpha \beta \gamma_0+\gamma_{12})(R_{se}\rho R_{sg}+R_{gs}\rho R_{es})\,, \\
\left(\frac{\partial{\rho}}{\partial t}\right)_a &= -\Gamma_a\{(R_{ee}+R_{aa})\rho+\rho(R_{ee}+R_{aa}) \nonumber \\
& -2(R_{ae}\rho R_{ea}+R_{ga}\rho R_{ag})\} \nonumber \\
& -(2\alpha \beta \gamma_0-\gamma_{12})(R_{ae}\rho R_{ag}+R_{ga}\rho R_{ea})\,,\\
\left(\frac{\partial{\rho}}{\partial t}\right)_I &= -\Gamma_{I}\{(R_{as}+R_{sa})\rho+\rho(R_{as}+R_{sa}) \nonumber \\
&-2(R_{ga}\rho R_{sg}+R_{gs}\rho R_{ag}+R_{se}\rho R_{ea}  \nonumber\\
& +R_{ae}\rho R_{es})\}  +(\alpha^2-\beta^2)\gamma_0\{R_{ae}\rho R_{sg}\nonumber\\
&+R_{gs}\rho R_{ea}+R_{se}\rho R_{ag}+R_{ga}\rho R_{es}\}\,,
\end{align}
\end{subequations}
with the damping coefficients
\begin{subequations}
\begin{align}
\Gamma_s &= \frac{1}{2}(\gamma_0+2\alpha \beta \gamma_{12})\,,\\
\Gamma_a &= \frac{1}{2}(\gamma_0-2\alpha \beta \gamma_{12})\,,\\
\Gamma_{I} &= \frac{1}{2}(\alpha^2-\beta^2)\gamma_{12}\,.
\end{align}
\end{subequations}
$\gamma_0,\gamma_{12}$ are the Einstein $A$ coefficient and the dipole-dipole cross damping rate, respectively. 
The collective qubit operators are defined as $R_{ij}=|i\rangle \langle j|$ where the collective states $|i\rangle, |j\rangle$  ($i,j\in\{e,g,s,a\}$) are given by Eq.~(\ref{eq:CBasis}).
The contribution with subindex $s$ [$a$] describes the spontaneous decay via the symmetric state $|s\rangle$ [anti-symmetric state $|a\rangle$]. The part with index $I$ is an interference part involving both the symmetric and the anti-symmetric states. It results from spontaneously induced coherences between the symmetric and antisymmetric transitions, and only contributes if $\alpha^2\neq \beta^2$.

The second incoherent contribution $\mathscr{L}_\Gamma \rho$ takes the form
\begin{equation}\label{eq:LrhoG}
\mathscr{L}_\Gamma \rho=\tilde\gamma_0(2R_{ge}\rho R_{eg}-R_{ee}\rho-\rho R_{ee})\,.
\end{equation}
It arises from the contributions proportional to $\chi_r^{(lm)}$ in Eq.~(\ref{eq:HB}). Note that the decay rate $\tilde\gamma_0$ is smaller than $\gamma_0$ according to Liu's work~\cite{YXLiuPRL96p067003}, but their ratio is tunable during the fabrication of a superconducting circuit. In order to simplify the discussion, we assume that $\tilde\gamma_0=\gamma_0$.

By applying TDMFs with different frequencies, selective processes described by  $H_Q$  become resonant, such that various types of effective interaction Hamiltonian can be generated.
In the following investigation, the qubits work at two slowly-varying frequencies $\omega_0^{(1)}$ and $\omega_0^{(2)}$, respectively, controlled by the bias fluxes $\Phi_e^{(l)}$. The frequency difference $\Delta$ is kept much smaller than the always-on coupling strength $\lambda_1$. As a consequence, the slowly-varying bias magnetic flux does not excite unwanted transitions between the symmetric and antisymmetric states.

Throughout our investigations below, we assume flux qubits with dephasing times $T_\varphi= 1\sim 10 \mu s$, as observed in recent experiments~\cite{JClarkeNature453p1031,JQYouPhysToday58p42,PBertetPRL95p257002}, and an energy relaxation time $T_{E}$ of roughly half the dephasing time $T_\varphi$. Thus, if the bias flux is changed in order to tune the transition frequencies, they vary in a range smaller than $60$~MHz.
Our numerical results show that the influence of this variation on the circulating current $I^{(l)}$, and subsequently on $\lambda_1$, is negligible. Experimentally, always-on coupling strengths $|\lambda_1|$ of several hundred MHz or even higher are realized~\cite{JBMajerPRL94p090501,AIzmalkovPRL93p037003,MGrajcarPRL96p047006}.  Note that the sign of $\lambda_1$ can be controlled by choosing ferromagnetic or anti-ferromagnetic coupling~\cite{MGrajcarPRL96p047006,JQYouPRB71p024532}. Therefore, it is possible to compensate the energy shift from the DDI via  $\lambda_1$.

\section{\label{sec:g2a}Preparation of the antisymmetric state}
We first aim at populating the so-called antisymmetric state of the two-qubit system. For this, the frequency of the applied TDMF is chosen close to the average frequency $\omega_0$. Then, the terms proportional to $\Omega_{lm}^{(1)}$, $\Omega_{lm}^{(2)}$ can be dropped in a rotating wave approximation (RWA) in Hamiltonian~(\ref{eq:HQ}).

We apply the master equation approach to the dynamics of the system in a frame rotating with the frequency of the TDMF $\omega_c$. In the collective states basis, the master equation for the system density matrix $\rho$ takes the form
\begin{equation}\label{eq:MEqg2a}
\dot \rho=\frac{i}{\hbar}[\rho,H_0+H_I]+\mathscr{L}\rho+\mathscr{L}_\Gamma\rho\,,
\end{equation}
where
\begin{subequations}
\begin{align}
H_0 =& \hbar[\delta (2R_{ee}+R_{ss}+R_{aa})+w(R_{ss}-R_{aa})]\,,
\label{eq:hambare}\\
H_I =& \hbar(\alpha+\beta) \left[ \Omega (R_{es}+R_{sg})+H.c.)\right]\nonumber\\
& +\hbar(\alpha-\beta) \left[\Omega (R_{ea}+R_{ag})+ H.c.\right]\,.
\end{align}
\end{subequations}
The detuning $\delta=\omega_0-\omega_c$ and the Rabi frequency $\Omega=-k_1=-k_2$, see Eq.~(\ref{eq:HQ}). We set the energy of level $|g\rangle$ to zero. The amplitude of $\Omega$ is $\Omega_0$, whereas the phase $\varphi$ of $\Omega$ does not influence the population, and thus is ignored in the following. The dissipative part $\mathscr{L}\rho+\mathscr{L}_\Gamma\rho$ is defined in Eqs.~(\ref{eq:Lrho}) and (\ref{eq:LrhoG}).
 
The antisymmetric state is of interest since it is partially decoupled from the interaction with the reservoir as $\Gamma_s < \gamma_0/2$ and therefore more stable against decoherence. However, as is well known from atomic systems, this decoupling at the same time makes a controlled population of this state difficult, as it also decouples the state from driving fields. This in particular holds if the two involved qubit transition frequencies are identical.

\begin{figure}[t]
\centering
\includegraphics[width=0.9\linewidth]{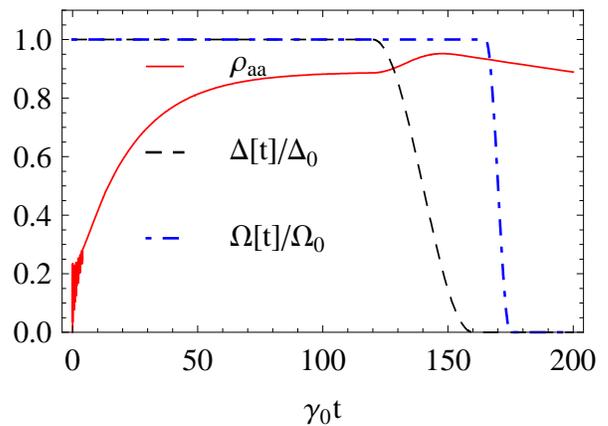}
\caption{(Color online) Time evolution of the population (solid red line) in the antisymmetric state $|a\rangle$. The idea is to prepare the anti-symmetric state while the two qubits are non-degenerate, and only afterwards render the two qubits degenerate. For this, the parameters are chosen such that $\gamma_{12}=0.9986\,\gamma_0$, $\lambda=50\,\gamma_0$, $\delta=50\,\gamma_0$, $\Omega_0=50\,\gamma_0$. The two qubit transition frequencies are adjusted via time-dependent bias fluxes, such that the frequency difference $\Delta (t)$ (dashed black line) changes from $18\gamma_0$  to zero as a cosine function during the time period $120\gamma_0^{-1}$ to $160\gamma_0^{-1}$. The driving field (dash-dotted blue line) is turned off from its initial value $\Omega_0$ in the period $165\gamma_0^{-1}$ to $175\gamma_0^{-1}$.}
\label{fig:figPadtn}
\end{figure}

But as an essential difference of the flux qubit system to atomic systems, the transition frequency of a flux qubit can be individually controlled by its bias flux~\cite{JMajerNature449p443,AIzmalkovPRL93p037003,MGrajcarPRL96p047006}. In the following, we exploit this feature and demonstrate that the antisymmetric state can be populated even for identical flux qubits by first adjusting the bias fluxes such that the qubits become unequal, then preparing the antisymmetric state, and finally switching back to the degenerate case. 

An example for this is shown in Fig.~\ref{fig:figPadtn}. Initially, the two qubits have a frequency difference $\Delta(t=0)=\Delta_0=18\gamma_0$. Applying a continuous TDMF $\Omega$ during $0\leq \gamma_0t\leq 165$ allows to populate the antisymmetric state, as can be seen in Fig.~\ref{fig:figPadtn}. After a certain time ($\gamma_0 t = 120$ in our example), the bias fluxes are continuously adjusted such that the two qubits become degenerate, $\Delta(\gamma_0 t\geq160) = 0$.  It can be seen from Fig.~\ref{fig:figPadtn} that a preparation fidelity for the antisymmetric state in the degenerate two-qubit system of about $F=0.94$ is achieved.  Finally, the TDMF is switched off as well in the time period $165\leq \gamma_0t\leq 175$, demonstrating that it is not required to preserve the population in the antisymmetric state.  It should be noted that this scheme does not rely on a delicate choice and control of parameters, as it is the case, e.g., for state preparation via special-area pulses.

The limited preparation fidelity and the slow decay of the state is due to the fact that the decoupling of the anti-symmetric state is not perfect. For any realistic system, the distance between the two qubits remains finite, such that $\Gamma_a >0$. For our parameters, $\gamma_{12}\approx0.9986\gamma_0$, and the decay rate of the antisymmetric state is found to  be $\Gamma_a \approx 0.0014\gamma_0$, strongly suppressed by the dipole-dipole coupling.
In this example, the energy shift on the antisymmetric and symmetric states induced by the direct DDI is of order $10^3\,\gamma_0$. As discussed before, it can be compensated via $\lambda_1$ to yield a relatively small $\lambda$. We found that the maximum fidelity obtained in our example is insensitive to $\lambda$ in a range of about $\lambda=40\gamma_0\sim100\gamma_0$.

Finally, we note that the antisymmetric state is also an entangled state. We use the concurrence $C$~\cite{JAudretschEntSys,SNataliPRA75p042307} as an entanglement measure, which is given by
\begin{subequations}
\begin {align}
C  =&2 \max\{0,\chi(t)\}\,,\\
\chi(t)=&\left|\alpha \beta (\rho_{ss}-\rho_{aa})+\alpha^2 \rho_{sa}-\beta^2 \rho_{as}\right|\nonumber \\
&-\sqrt{\rho_{ee} \rho_{gg}}\,.
\end{align}
\end{subequations}
In our example, $C$ approaches $0.89$ at time $t=160\gamma_0^{-1}$.


\section{\label{sec:a2s}Preparation of the symmetric state}
In this section, we discuss the preparation of the symmetric entangled state $|s\rangle$. Three approaches are compared.
First, we directly prepare the symmetric state from the ground state using a special-area pulse. Second, SCRAP is used to populate the symmetric state from the ground state. Finally, $|s\rangle$ is populated via the anti-symmetric state discussed in the previous section~\ref{sec:g2a}.

\begin{figure}[t]
\centering
\subfigure{\label{subfig:figg2sP}
\includegraphics[width=0.9\linewidth]{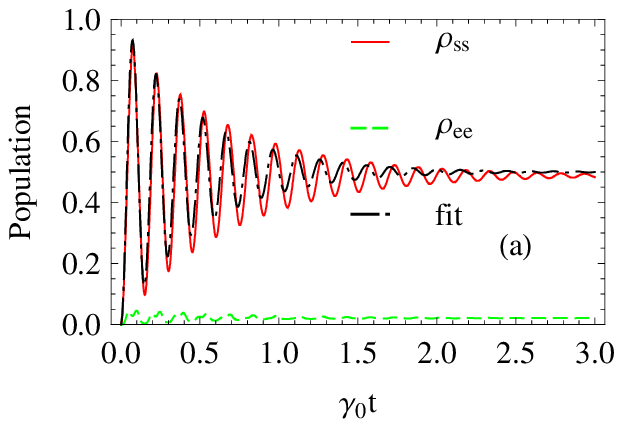}}\\
\subfigure{\label{subfig:figg2sC}
\includegraphics[width=0.9\linewidth]{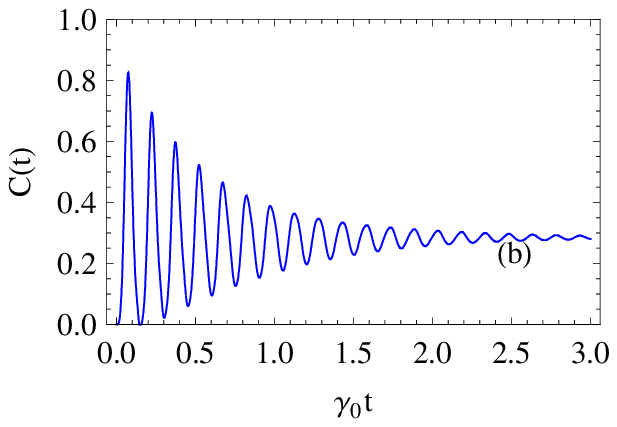}}\\
\caption{\label{fig:figg2s}(Color online) (a) Time dependent population of the symmetric state $|s\rangle$ directly from the ground state. The dynamics is induced by a continuous TDMF with parameters $\Omega_0=15\gamma_0$ and $\delta=-50\gamma_0$, and the other parameters are $\gamma_{12}=0.9986\,\gamma_0$, $\lambda=50\,\gamma_0$, and $\Delta=0$. The numerical result (solid red line) is well fitted by Eq.~(\ref{eqn-ss-symm}) shown as the black dash-dotted line. A small part of the population is transfered to the collective excited state (dashed green line). (b) Corresponding results for the concurrence.}
\end{figure}

\subsection{\label{sec:direct}Direct population via special-area pulse}
First, we populate the symmetric state directly from the ground state. If the two qubits are closely-spaced,  the dipole-dipole  level shift is large enough for an almost selective excitation of the collective states without populating the collective excited state $|e\rangle$. For example, it was shown that in such a system of two identical two-level atoms, the selective population of $|s\rangle$  is possible through a short $\pi$-area pulse, even though there are competing channels such as exciting the atoms from the symmetric channel to the collective excited state~\cite{ABeigeJMO47p401}. In the following, we also consider two identical flux qubits with equal transition frequencies, and assume that they experience the same driving TDMF. The TDMF is chosen to be near resonant to the $|g\rangle \leftrightarrow |s\rangle$ transition, i.e., $\delta\approx-\lambda$.

In analogy to Ref.~[\onlinecite{ABeigeJMO47p401}], in our scheme described by Eq.~(\ref{eq:MEqg2a}), the state $|s\rangle$ can also be selectively prepared by a standard $\pi$-area TDMF pulse with sufficiently large detuning to the $|s\rangle \leftrightarrow |e\rangle$ transition. When the system  initially is in its ground state, the maximum population of state $|s\rangle$ is obtained at time $t=\pi/(2\sqrt{2}\Omega_0)$.

In Fig.~\ref{subfig:figg2sP}, we show results for the population of $|s\rangle$ for continuous driving. It can be seen that the population reaches a maximum value, but afterwards exhibits rapid oscillations at frequency $2\sqrt{2}\Omega_0$, while the amplitude of the subsequent maxima in the population decays as an exponential function $\exp[-(\gamma_0+\gamma_{12}) t]$ until the system approaches its stationary state.
This result can be understood by reducing the system to a two-state system only involving in the states $|s\rangle$ and $|g\rangle$. The numerical results can be well fitted by the solution of this two-state approximation,
\begin{align}
\label{eqn-ss-symm}
\rho_{ss}^{\textrm{2-level}}(t) =& \frac{1-e^{-(\gamma_0+\gamma_{12}) t}}{2} \nonumber \\
&+e^{-(\gamma_0+\gamma_{12}) t} \sin^2 (\sqrt{2}\Omega_0 t)\,,
\end{align}
as can be seen from Fig.~\ref{subfig:figg2sP}.
The time evolution of the concurrence shown in Fig.~\ref{subfig:figg2sC} exhibits oscillations along with the population of $|s\rangle$. As expected, the maximum concurrence of $C=0.83$ occurs at time $\pi/(2\sqrt{2}\Omega_0)$.

We chose to display the result for a continuous driving field rather than for a $\pi$-pulse in order to illustrate that the maximum population of state $|s\rangle$ is strongly dependent on the parameters of the driving TDMF. If the Rabi frequency $\Omega_0$ or the detuning $\delta$ are not precisely controlled in an experiment, then the optimum $\pi$-area pulse is not applied, and the population of the excited state is strongly reduced. In our example, there is an optimum Rabi frequency of about $15\gamma_0$. For resonant excitation $(\delta=-\lambda)$ with this Rabi frequency, the population of state $|s\rangle$ reaches its maximum value $0.90$ at time $0.07\gamma_0^{-1}$.
For $\Omega_0=5\gamma_0$, the maximum population decreases to $0.84$, while the maximum concurrence becomes to $0.79$. For $\Omega_0=25\gamma_0$, the maximum population and concurrence are only $0.78$ and $0.60$, respectively. For a non-ideal detuning
$\delta=-40\gamma_0$, the population and the concurrence have maximum values of $0.80$ and $0.64$, respectively.

\subsection{\label{sec:direct-scrap}Direct population via SCRAP}

So far, as in our previous section, in most cases special-area pulses have been used to create coherent superpositions in SQs~\cite{RMcdermottScience307p1299,YXLiuPRL96p067003,JMajerNature449p443,MASillanpaaNature449p438}.
 However as discussed above, this technique is not robust: variations in pulse area and detuning from resonance can lead to considerable loss in preparation fidelity~\cite{JClarkeNature453p1031}. In order to overcome these problems, very recently, a so-called Stark-chirped rapid adiabatic passage (SCRAP) scheme for robust population transfer known from atomic systems~\cite{coherent-passage} has been proposed in superconducting current-biased qubits~\cite{LFWeiPRL100p113601}.

The SCRAP technique builds on the well-known method of rapid adiabatic passage and works as follows. First, a pump laser pulse  tuned slightly away from resonance with the transition between two bound states is applied, and then a second delayed Stark pulse sweeps the bound states through the resonance by inducing a dynamic Stark shift. Throughout this process, the population is adiabatically transferred between the two states. Also arbitrary superpositions of two bound states can be achieved via the SCRAP technique~\cite{LPYatsenkoOC204p413}.

SCRAP is particularly useful in our system, since a modification of the bias fluxes allows for a convenient control of the qubit transition frequencies. To apply a SCRAP scheme to populate the symmetric state, we choose Gaussian shaped pump TDMF and bias fluxes,
\begin{subequations}\label{eq:SCRAP}
\begin{align}
\Omega(t) &=\Omega_0 e^{-(t-\tau_p)^2/T_p^2}\,,\\
\delta(t) &=\delta_0-S_0 e^{-(t-\tau_s)^2/T_s^2}\,,
\end{align}
\end{subequations}
with $\Omega_0=32\,\gamma_0$, $S_0=18\,\gamma_0$, $T_s=0.02\,\gamma_0^{-1}$, $T_p=T_s$, $\tau_s-\tau_p=T_s$ and $\delta_0=-60\,\gamma_0$ corresponding to an exact detuning $-10\,\gamma_0$.
The delayed dynamic bias flux follows the pump pulse. For this, the position of the symmetric collective state is adjusted in time via the bias flux.
\begin{figure}
\centering
\includegraphics[width=0.9\linewidth]{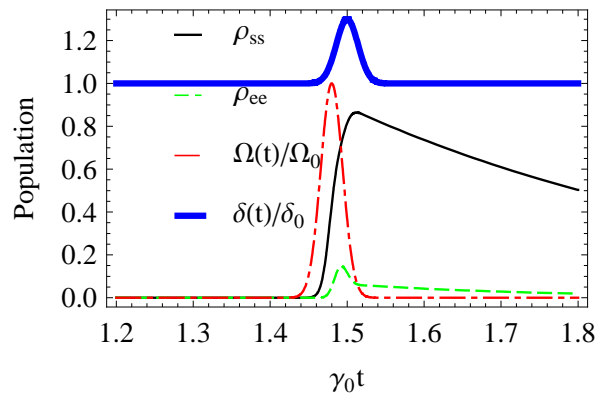}\\
\caption{\label{fig:SCRAPg2s}(Color online) Robust populating the symmetric state from the ground state via the SCRAP technique for $\gamma_{12}=0.9986\, \gamma_0$, and $\lambda=50\, \gamma_0$. The solid black line shows the population of the desired symmetric state, while the thick solid blue line and the dash-dotted red line are the time-dependent Rabi frequency and detuning required for SCRAP.}
\end{figure}
Because the competing channel $|s\rangle\leftrightarrow |e\rangle$ limits the intensity and duration of the pump pulse, the population (black line in Fig.~\ref{fig:SCRAPg2s}) in the symmetric state is limited to about $0.81$ directly after the SCRAP preparation at time $1.55\, \gamma_0^{-1}$. The corresponding concurrence is $0.64$. Even though this fidelity is rather low compared to the other preparation schemes presented here, the SCRAP approach may be an alternative if suitable and reliable special-area pulses to populate $|s\rangle$ are hard to achieve experimentally.

\subsection{Population via the anti-symmetric state}

In Sec.~\ref{sec:g2a} we have shown that a nearly complete populating of state $|a\rangle$ can be obtained for two identical flux qubits, without the need for precisely controlled TDMF pulses. In this subsection, we extend this scheme by preparing the symmetric state from the antisymmetric state in two identical flux qubits. In comparison with the direct preparation of $|s\rangle$ from the ground state discussed in Sec.~\ref{sec:direct}, the population is first transferred to $|a\rangle$ and afterwards transferred to $|s\rangle$. If the driving field is applied continuously, then the system will turn out to mainly oscillate between the two maximally entangled states $|a\rangle$ and $|s\rangle$. In contrast to simple special-area pulse schemes, the two Rabi frequencies driving the qubits are applied with a fixed relative phase, which is possible since we work around the optimum point, as discussed below.

To selectively excite the transition between state $|a\rangle$ and $|s\rangle$ under the condition of $\lambda \gg|\Delta|$, the frequency $\omega_c$ of the TDMF should be close to the level shift $2\lambda$ rather than to the frequency difference $\Delta$. In this case, a rotating wave approximation eliminates terms proportional to $k_l$ and $\Omega_{lm}^{(2)}$ in the Hamiltonian Eq.~(\ref{eq:HQ}), such that the contribution proportional to $\Omega_{lm}^{(1)}$ determines the dynamics. The master equation then reads
\begin{align}
\label{eq:Mas}
    \frac{\partial \rho}{\partial t}& =-i\left[\delta
    (R_{ss}-R_{aa}),\rho \right] \nonumber \\
    & \quad  +i\left[\Omega R_{as}+\Omega^* R_{sa},\rho
     \right]+\mathscr{L}\rho+\mathscr{L}_\Gamma \rho,
\end{align}
where the detuning $\delta=w-\omega_c /2$. The Rabi frequency is $\Omega=\alpha^2\zeta-\beta^2 \zeta^*=(\alpha^2 -\beta^2)\Re[\zeta]+i \Im[\zeta]$ with $\zeta=\Omega_{12}^{(1)}+\Omega_{21}^{(1)*}$. Here, $\Re[~]$ and $\Im[~]$ denote the real part and the imaginary part, respectively. Since $\alpha^2-\beta^2 \approx 0$ for small $\Delta$, a large imaginary part of $\Omega$ is required to efficiently drive the system. Thus, the phase of the driving TDMF should be close to $\pi/2$, i.e., $\Omega=i\Omega_0$. This can be achieved by setting the phases $\theta_{12}^{(1)}$ and $\theta_{21}^{(1)}$ of the Rabi frequencies $\Omega_{12}^{(1)}$ and $\Omega_{21}^{(1)}$ to $\pi/2$ and $3\pi/2$, respectively~\cite{YXLiuPRL96p067003}.
We have numerically solved the Schr\"odinger equation  to evaluate $\Omega_{12}^{(1)}$ and $\Omega_{21}^{(1)}$. We found that the two coupling coefficients have the same magnitude but opposite signs if the two bias fluxes are at symmetric positions with respect to the optimal point $f_{l}=0.5$. Therefore, it is possible to individually control the phases of the Rabi frequencies via the bias fluxes as well as the phases of the driving TDMF.

In this configuration, the driving TDMF is well off-resonant from the transitions of $|g\rangle\leftrightarrow |s\rangle$ and $|e\rangle\leftrightarrow |s\rangle$, such that no such transitions are induced by the TDMF. Still, the population is spontaneously damped from the symmetric state $|s\rangle$ to the ground state. The equation of  motion for the population in the ground state is obtained as
\begin{align}
 \frac{\partial \rho_{gg}}{\partial t}=\gamma_0\,
[1-\rho_{gg}(0)]+2\alpha \beta\, \gamma_{12}\,(\rho_{ss}-\rho_{aa}).
\end{align}
This can be solved to give
\begin{align}
\rho_{gg}(t)&=1-[1-\rho_{gg}(0)]\, e^{-\gamma_0 t}\nonumber\\
&\quad +\int_{0}^{t}{ 2\,\alpha \beta\, \gamma_{12}\,[\rho_{ss}(t')-\rho_{aa}(t')]\, dt'}\nonumber\\
&\sim 1-[1-\rho_{gg}(0)]\, e^{-\gamma_0 t}\,.
\end{align}
\begin{figure}[t]
\centering
\subfigure{\label{subfig:a2s15P}
\includegraphics[width=0.9\linewidth]{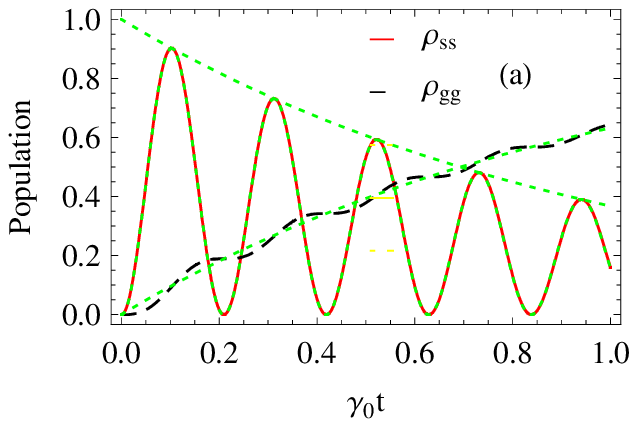}}\\
\subfigure{\label{subfig:a2s15C}
\includegraphics[width=0.9\linewidth]{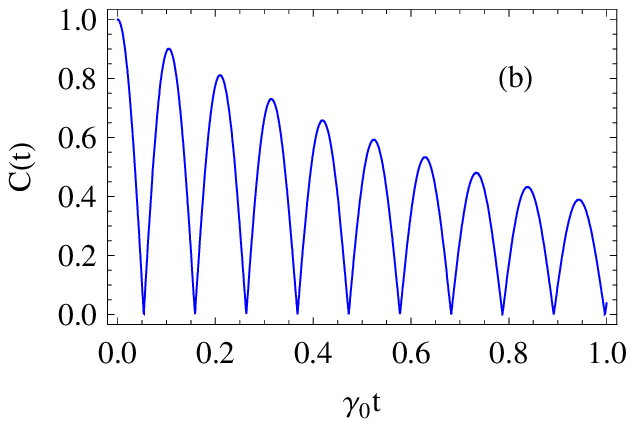}}\\
\caption{\label{fig:a2s15}(Color online) Population of the symmetric state $|s\rangle$ from the anti-symmetric state $|a\rangle$. The two panels show the time-evolution of (a) population and (b) concurrence for $\gamma_{12}=0.9986\gamma_0, \lambda=50\gamma_0, \Delta=0, \delta=0$ and $\Omega_0=15\gamma_0$. The populations in the state $|g\rangle$ (Black dashed line) and $|s\rangle$ (solid red line) are well fitted by $\rho_{gg}(t)$ and $\rho_{ss}(t)$ (fits shown by green dotted lines), respectively.}
\end{figure}
Here we drop the integral term which describes a small and rapidly oscillating perturbation. Similarly, the time-dependent population in the state $|s\rangle$ is
\begin{align}
\label{eqn-ss}
\rho_{ss}(t)=\frac{\rho_{aa}(0)\, \Omega_0}
{\sqrt{\delta^2+\Omega_0^2}}\, e^{-\gamma_0 t}\, \sin \left (\sqrt{\delta^2+\Omega_0^2}\,t\right )\,.
\end{align}
In the above equations, $\rho_{gg}(0)$ and $\rho_{aa}(0)$ denote the initial populations in the states $|g\rangle$ and $|a\rangle$, respectively.

To study the fidelity of the population transfer between $|a\rangle$ and $|s\rangle$, in Fig.~\ref{subfig:a2s15P}, we choose the anti-symmetric state as the initial condition. Applying a continuous-wave TDMF with Rabi frequency $\Omega_0=15\,\gamma_0$ and detuning $\delta=0$, the symmetric state $|s\rangle$ reaches its maximum population of $0.90$ at time $0.1\,\gamma_0^{-1}$. After this maximum, the population continues to oscillate between $|a\rangle$ and $|s\rangle$ due to the applied field. This oscillation is damped by an overall decay as we include damping with a rate $\gamma_0$. The corresponding concurrence is shown in in Fig.~\ref{subfig:a2s15C}. The concurrence oscillates at twice the frequency of the population oscillation, since both $|s\rangle$ and $|a\rangle$ are maximally entangled. The local maximum values of the entanglement occur at times $n\pi/(2\sqrt{\delta^2+\Omega_0^2})$ where either $|s\rangle$ or $|a\rangle$ is occupied.

One can improve the maximum population transfered to $|s\rangle$ by increasing the Rabi frequency, because then the transfer to the symmetric state is more rapid and thus leads to less damping throughout the transfer. This enhancement is limited by the fact that the Rabi frequency must not become strong enough to also induce transitions involving the collective excited or ground states or to break the rotating wave approximation. For example, a population of the symmetric state of $0.97$ is obtained if we use a TDMF with $\Omega_0=50\, \gamma_0$ and $\delta=0$ with duration $\pi/(2\Omega_0)$, i.e., a $\pi$-pulse. In this case, a concurrence of $0.97$ is achieved.


We also found that the population transfer is rather insensitive to the detuning. For example, with $\Omega_0=50\,\gamma_0$ and $\delta=7\gamma_0$, the population of the symmetric state changes only slightly. Similarly, the scheme works well if only parts of the population are initially in the state $|a\rangle$, as can be seen from Eq.~(\ref{eqn-ss}).

\section{\label{sec:g2e}Preparation of Bell states and the collective excited state}
Interacting atomic systems do not allow for an efficient transfer of population
to the collective excited state, as the transfer always has to proceed via either the collective symmetric or the antisymmetric intermediate state. In the following, we show that a controlled change of the flux qubit properties allows
to robustly generate the collective excited state, as well as the Bell states
\begin{align}
|\Phi_{\pm}\rangle=\frac{1}{\sqrt{2}}(|g_1,g_2\rangle\pm |e_1,e_2\rangle)\,,
\end{align}
via a mechanism not present in atomic systems. For this, we now assume the frequency-matching condition $\omega_c=\omega_0^{(1)}+\omega_0^{(2)}$ in Eq.~(\ref{eq:HQ}).

\begin{figure}[t]
\centering
\includegraphics[width=0.9\linewidth]{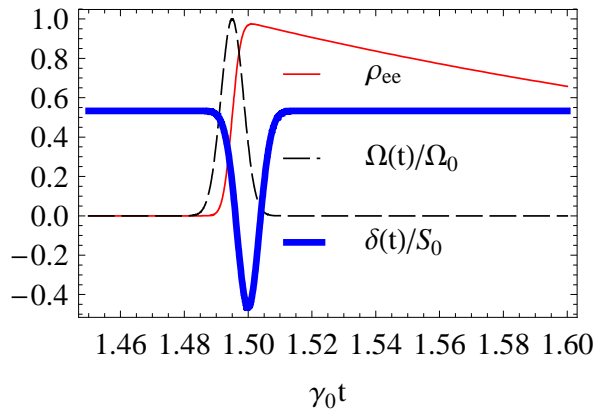}\\
\caption{\label{fig:SCRAPg2edmp}(Color online) Direct population transfer from the collective ground to the collective excited state (solid red line) using the SCRAP technique. A Gaussian pump pulse (dashed black line)
is followed by a delayed Gaussian time-dependent detuning (blue thick line).}
\end{figure}

Under these conditions, a process described by the term proportional to $\Omega_{lm}^{(2)}$ in Eq.~(\ref{eq:HQ}) is possible that does not occur in atomic systems, where both qubits evolve
to their excited states after absorbing one common photon. The equation of motion for the density matrix becomes
\begin{align}
\dot\rho=\frac{i}{\hbar}[\rho,H]+\mathscr{L}\rho+\mathscr{L}_\Gamma \rho\,,
\end{align}
with
\begin{subequations}
\begin{align}
H=&\hbar\delta[2R_{ee}+R_{ss}+R_{aa}]+\hbar w[R_{ss}-R_{aa}]\nonumber\\
&-\hbar[\Omega R_{eg}+\Omega^*R_{ge}]\,,
\end{align}
\end{subequations}
where the detuning $\delta=\omega_0-\omega_c/2$ and $\Omega$ is the Rabi frequency. Interestingly, we find that this system consisting of two coupled flux qubits behaves similarly to a two-level system composed of the collective ground state $|g\rangle$ and the collective excited state $|e\rangle$. Our numerical results show that this system is modifications of the parameters $\Delta$, $\lambda$ and $\gamma_{12}$. Therefore, in the following, we use fixed parameters $\Delta=0$, $\lambda=50\,\gamma_0$ and $\gamma_{12}=0.9986\,\gamma_0$ for simplicity.

\begin{figure}[tb]
\centering
\includegraphics[width=0.9\linewidth]{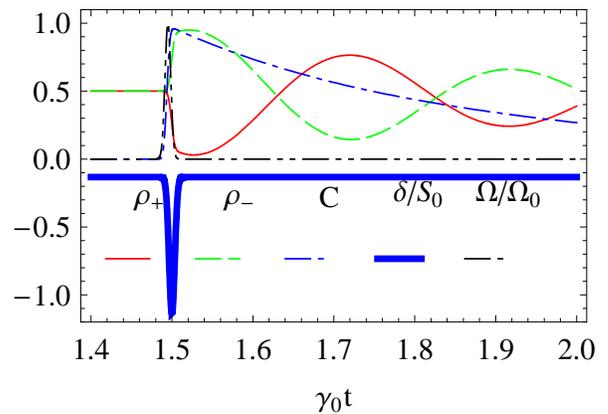}\\
\caption{\label{fig:BellSCRAP}(Color online) Creation of Bell states $|\Phi_\pm\rangle$ using the SCRAP technique from the ground state. The solid red line denotes population $\rho_+$ in $|\Phi_+\rangle$, while the dashed green line shows population $\rho_-$ in $|\Phi_-\rangle$.  The concurrence (dash-dotted blue line) has a maximum value of $0.94$. The black dash double dotted line indicates the applied SCRAP pulse Rabi frequency, while the thick blue line shows the time-dependent Stark detuning. }
\end{figure}

In order to achieve robust control, we use again the SCRAP technique to efficiently transfer population to the collective excited state $|e\rangle$ and to create superpositions of the collective ground and excited states. In addition to previous work~\cite{LFWeiPRL100p113601}, we study two coupled qubits, and include decoherence.

In Fig.~\ref{fig:SCRAPg2edmp}, a pump TDMF (black line) transfers the population to the collective excited state during the first resonant crossing. The dynamic detuning (blue thick line) is achieved by slowly tuning the bias flux.
To suppress the effect of the relaxation during the preparation, the pump laser pulse and the dynamic variation of the bias flux pulse ideally should be much shorter than the collective relaxation time $(4\gamma_0)^{-1}$.
We also choose Gaussian shaped pump TDMF and bias fluxes given in Eq.~(\ref{eq:SCRAP}) but apply different parameters: $\Omega_0=180\,\gamma_0$, $\delta_0=16\,\gamma_0$, $S_0=30\,\gamma_0$, $T_s=0.005\,\gamma_0^{-1}$, $T_p=T_s$, $\tau_s-\tau_p=T_s$. In our example, directly after the chirping of the detuning at time $1.51\gamma_0^{-1}$, the population of state $|e\rangle$ is $0.94$. Similarly, a so-called half-SCRAP with a small static detuning
$\delta_0 \approx 0$ can be used to robustly create  an arbitrary superposition of states $|e\rangle$ and $|g\rangle$ in principle~\cite{LPYatsenkoOC204p413}.

\begin{figure}[tb]
\centering
\includegraphics[width=0.9\linewidth]{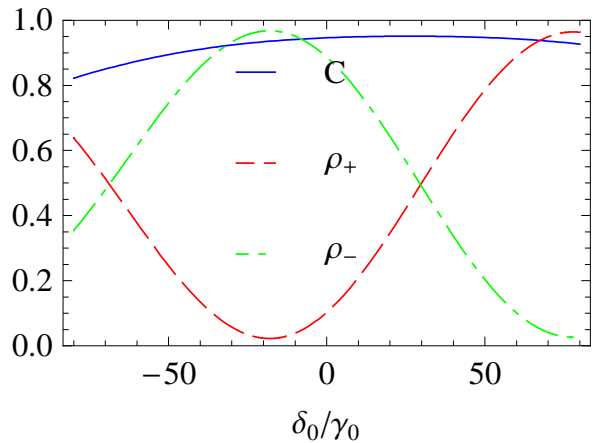}\\
\caption{\label{fig:BellSCRAPdtn}(Color online) Creation of superpositions of Bell states controlled by the static detuning $\delta_0$. The populations $\rho_\pm$ in states $|\Phi_+\rangle$ (dashed red line) and $|\Phi_-\rangle$ (dash-dotted green line) exhibit periodic oscillations as a function of $\delta_0$. The maximum concurrence $C$ is larger than $0.95$ (solid blue line).}
\end{figure}

The SCRAP technique can also be used to generate Bell states $|\Phi_{\pm}\rangle$ consisting of the collective ground and
excited states. An example is shown in Fig.~\ref{fig:BellSCRAP}. For the chosen parameters $\Omega_0=80\,\gamma_0$, $\delta_0=-8\,\gamma_0$, $S_0=60\,\gamma_0$, $T_s=0.005\,\gamma_0^{-1}$, $T_p=T_s,\tau_s-\tau_p=T_s$, the fidelity of the Bell state $|\Phi_-\rangle$ is $0.94$ at time $1.51\gamma_0^{-1}$.
Since $|\Phi_+\rangle$ and $|\Phi_-\rangle$ are not eigenstates of the coupled system~\cite{JMajerNature449p443}, oscillations occur
between these two states. At the same time, the oscillation amplitude is damped due to the decay from $|e\rangle$ to $|g\rangle$.
The concurrence is calculated in this case as~\cite{JAudretschEntSys}
\begin{subequations}
\begin{align}
C(t)&=2 \max\{0,|\rho_{eg}|-\sqrt{\rho_{22}\rho_{33}}\}\,,\\
\rho_{22}&=\beta^2 \rho_{ss}+\alpha\beta (\rho_{as}+\rho_{sa})+\alpha^2 \rho_{aa}\,,\\
\rho_{33}&=\alpha^2 \rho_{ss}-\alpha\beta (\rho_{as}+\rho_{sa})+\beta^2 \rho_{aa}\,,
\end{align}
\end{subequations}
where $\rho_{22}$ and $\rho_{33}$ are the populations of the product states $|e_1g_2\rangle$ and $|g_1e_2\rangle$.
The concurrence decreases together with the populations of the Bell states exponentially from an initial maximum value of $0.94$. Thus we find that Bell states $|\Phi_-\rangle$ can be created with high fidelity by means of the SCRAP technique.

Finally, we show that controlled superpositions of the two Bell states $|\Phi_\pm\rangle$ can be prepared using our scheme. For this, we focus on the state of the system directly after the SCRAP preparation at time $1.51\gamma_0^{-1}$. The concurrence and the fidelity of the Bell states $|\Phi_\pm\rangle$ as a function of the static detuning $\delta_0$ are shown in Fig.~\ref{fig:BellSCRAPdtn}. It can be seen that one can generate arbitrary superpositions of these two Bell states $|\Phi_\pm\rangle$ with corresponding probabilities varying between $\sim0.03$ and $\sim0.96$ in combination with a near complete entanglement. Because a very small part of the population is in the symmetric and anti-symmetric states, the sum population in two Bell states only $0.99$, slightly smaller than unity. The fidelity of $|\Phi_\pm\rangle$ sinusoidally oscillates with $\delta_0$ over a period of about $200\gamma_0$. It should be noted that the high maximum fidelity of $0.96$ is achieved although damping processes are included.

\section{\label{sec:conclusion}Conclusion}

In summary, robust schemes to create entangled states or controlled superpositions of entangled states in a system of two coupled flux qubits have been discussed. Decoherence of the qubits is included via the interaction of the system with a reservoir of harmonic oscillators. First, we have shown how the antisymmetric collective state can be prepared in two identical flux qubits in the steady state using continuous-wave driving fields. Since the antisymmetric state decouples from other states, it is useful, e.g., for the study of Bell's inequality violations~\cite{JClarkeNature453p1031}. We have also compared two different channels, direct and via the anti-symmetric state, to populate the symmetric collective state. Finally, a robust and flexible SCRAP technique has been demonstrated to efficiently
populate the collective excited state.  This technique enables one to create arbitrary superpositions of Bell states $|\Phi_\pm\rangle$ consisting of the collective ground and excited state simply by controlling the detuning.

\end{document}